\documentclass[10pt]{iopart}
\usepackage{graphicx}
\newcommand {\be}{\begin{equation}}
\newcommand {\ee}{\end{equation}}

\begin{document}

\paper[]{Coupled transport in rotor models}
\date{\today}

\author{S Iubini$^{1}$, S Lepri $^{2,5}$, R Livi$^{3,4,5}$ and A Politi $^{6}$\
}
\address{$^{1}$Centre de Biophysique Mol\'eculaire (CBM), CNRS-UPR 4301 Rue Charles Sadron, F-45071 Orl\'eans, France}

\address{$^{2}$ Consiglio Nazionale
delle Ricerche, Istituto dei Sistemi Complessi, 
via Madonna del Piano 10, I-50019 Sesto Fiorentino, Italy
}

\address{$^{3}$ Dipartimento di Fisica e Astronomia, Universit\`a di Firenze, 
via G. Sansone 1 I-50019, Sesto Fiorentino, Italy 
}
\address{$^{4}$ Centro Interdipartimentale per lo Studio delle Dinamiche Complesse,
Universit\`a di Firenze.
}
\address{$^{5}$Istituto Nazionale di Fisica Nucleare, Sezione di Firenze,  
via G. Sansone 1 I-50019, Sesto Fiorentino, Italy}

\address{$^{6}$  Institute for Complex Systems and Mathematical Biology \& SUPA
University of Aberdeen, Aberdeen AB24 3UE, United Kingdom 
}
\ead{stefano.iubini@cnrs-orleans.fr,stefano.lepri@isc.cnr.it,\newline 
roberto.livi@unifi.it, a.politi@abdn.ac.uk}

\begin{abstract}
Steady non-equilibrium states are investigated in a one-dimensional setup in
the presence of two thermodynamic currents.
Two paradigmatic nonlinear oscillators models are investigated: an XY chain  and the
discrete nonlinear Schr\"odinger equation. Their distinctive feature is that the
relevant variable is an angle in both cases.
We point out the importance of clearly distinguishing between energy and heat flux. 
In fact, even in the presence of
a vanishing Seebeck coefficient, a coupling between (angular) momentum and energy
arises, mediated by the unavoidable presence of a \textit{coherent} energy flux.
Such a contribution is the result of the ``advection'' induced by the position-dependent 
angular velocity.
As a result, in the XY model, the knowledge of the two diagonal elements of the 
Onsager matrix suffices to reconstruct its transport properties.
The analysis of the nonequilibrium steady states finally allows to strengthen the 
connection between the two models.
\end{abstract}




\section{Introduction}
 
The physics of open (classical or quantum) many-particle systems is 
a vast interdisciplinary field ranging from the more basic  
theoretical foundations to the development of novel technological 
principles for energy and information management.
Within this broad context, simple models of classical nonlinear oscillators have 
been investigated to gain a deeper understanding of heat transfer processes far 
from equilibrium \cite{LLP03,DHARREV,basile08} (see also \cite{Lepri2016} for 
a recent account). 
The existing literature mostly focused on the case where just one
quantity, the energy, is exchanged with external reservoirs and transported across the system --
see e.g. \cite{Wang2013,Wang2011,Lepri2009,Das2014a,Lee-Dadswell2015} for some recent work.
In general, however, the dynamics of physical systems is characterized by more than one conserved quantity
each associated with a hydrodynamic mode of spontaneous fluctuations \cite{Spohn2014, Das2014,Liu2014}. 
Under the action of external reservoirs, one expects the corresponding currents to be coupled in the 
usual sense of linear irreversible thermodynamics.
A well-known example is that of thermoelectric phenomena whereby useful electric work can be extracted 
in the presence of temperature gradients. 

From the point of view of statistical mechanics, a few works have been so far devoted to coupled
transport: they can be grouped in those devoted to interacting particle 
gases \cite{Mejia2001,Casati2009,Benenti2014} and to coupled oscillator 
systems \cite{Gillan85,Basko2011,Iubini2012,DeRoeck2013,Borlenghi2015}.
The connection between microscopic interactions and macroscopic thermodynamic properties
is still largely unexplored.
In this paper, we provide a detailed characterization of coupled transport in possibly
the simplest dynamical model, the one-dimensional rotor model, also termed Hamiltonian 
XY model ~\cite{Escande1994}. Here, there are two conserved quantities (energy and
angular momentum), two associate currents, 
and only one relevant thermodynamic parameter, the temperature.

The simplicity of the model reveals the crucial role played by the \textit{coherent} energy flux,
normally present in steady nonequilibrium states: it represents the part of 
the energy current advected by the local average angular momentum.
In a sense, it is the mediator in the coupling 
between the two currents. As a result, it is absolutely necessary to distinguish
between energy and heat fluxes, as only the former one takes fully into account the
coherent contribution.

Previous studies of the Hamiltonian XY model (referred to in the following as XY model
for brevity) essentially focused on the transport of heat, in the absence of an
angular-momentum flux. In such a setup, the model is an example where transport
is normal in 1D in spite of the momentum being conserved
~\cite{Giardina99,Gendelman2000,Yang2005,Li2015}.
There are two complementary views to account for this behavior. In the general
perspective of nonlinear fluctuating hydrodynamics \cite{Spohn2014,Mendl2013}
normal diffusion can be 
explained by observing that the angle variables do not constitute a conserved field, 
which leads to the absence of long-wavelength currents in the system \cite{Das2014b}. 
From a dynamical point of view, one can invoke that normal transport sets in due to the
spontaneous formation of local excitations, termed
rotobreathers, that act as scattering centers \cite{Flach2003}. 
Phase slips (jumps over the energy barrier), on their side,  may  effectively
act as localized random kicks, that contribute to scatter the 
low-frequency modes, thus leading to a finite conductivity.
Actually, such long-lived localized structures lead also to anomalously
slow relaxation to equilibrium ~\cite{Eleftheriou2005,Cuneo2015}. 
Non stationary (time dependent) heat exchange processes have also been 
shown to be peculiar \cite{Gendelman2010}. 
The effect of external forces has been previously addressed 
only in Ref.~\cite{Iacobucci2011} and boundary-induced transitions have also 
been discovered \cite{Iubini2014} (see also \cite{Ke2014}).
The important extension to 2D is
characterized by the presence of a Kosterlitz-Thouless-Berezinskii
phase transition between a disordered
high--temperature phase and a low--temperature one, displaying anomalous
and normal transport respectively \cite{Delfini2005}.

More recently, the 1D XY model has attracted the interest in a different context
for some nontrivial properties related to the transport of angular momentum or, using
a different language, electric charge \cite{Pino2016}.
In fact, it can be also interpreted as the classical limit of an array of
Josephson junctions. In the quantum version, a many body localization phenomenon,
associated to an ergodicity breaking mechanism, has been observed
and proved to exist. In the classical limit,
the frequency can be interpreted as a charge variable, so that the transport of charge is
nothing but the current of angular momentum in the standard representation.

In Section \ref{theory} we review the general thermodynamic formalism of 
linear--response and then develop specific relationships for the XY model that 
are later used to interpret the results of numerical simulations. 

A careful analysis of nonequilibrium stationary states in coupled transport
requires an appropriate definition of the reservoirs controlling two fluxes at 
the same time. This point is discussed in Section \ref{thbath}, where we provide a comparison between a Langevin and a collisional stochastic scheme.

The results of numerical simulations of coupled transport in the
XY model are presented in Section \ref{numsim}, where
we also describe how to determine the dependence of the Onsager coefficients
on the temperature, when a suitable reference frame for the frequencies is adopted.
The numerical analysis confirms the prediction of linear Onsager theory, according
to which the Onsager coefficients of the XY model do not depend on the frequency and that
no coupled transport is present in the heat-representation.

In order to test to what extent the scenario reconstructed for the XY model 
applies to more general models, where thermodynamic properties depend also on 
the chemical potential, we study the Discrete NonLinear Schr\"odinger (DNLS) model 
and compare its nonequilibrium behavior with that of a 1D XY chain.
It was recently argued that the high mass-density regime
of the DNLS equation can be mapped onto an XY chain \cite{Iubini2013a}. 
In Section \ref{dnls} we reconsider the mapping between these two models 
in the framework investigating the corresponding Onsager coefficients.
As a result, we confirm the existence of a zero-Seebeck coefficient line,
whose very existence can be used as a reference to quantify the
deviations from the XY dynamics. 
Conclusions and perspectives are discussed in Section \ref{conclu}.
 
\section{Theoretical framework and the rotor chain model}
\label{theory}
A great deal of the recent literature on transport phenomena in one-dimensional
systems is focused on heat transport alone \cite{LLP03,DHARREV}.
In such cases, the relevant  physical observables are the heat flux $j_q$ 
and the corresponding  thermodynamic force, namely the gradient of temperature $T$  
(in what follows we equivalently refer to $T$ or  $\beta = 1/T$, selecting the more appropriate
quantity for the theoretical description). 
They are related by the Fourier equation
 \[
j_q =  - \kappa \,\, \frac{ d T}{dy} \; ,
\]
where $\kappa$ is the heat conductivity, and $y$ is the spatial direction of 
the applied gradient. The variable $y$ represents the spatial position along the chain (without prejudice of generality its length can be normalized to unit, i.e. $0\le y\le 1$)

In this section we describe the  formalism of coupled transport in one-dimensional
systems where a second quantity is transported: we call it ``momentum", but it could be 
any other physical observable like mass,  charge, etc. 
Its flux is denoted by $j_p$ and the corresponding thermodynamic force is the
gradient of chemical potential $\mu$. 

Within linear nonequilibrium thermodynamics, coupled transport can be characterized by making use of two equivalent representations. The heat--representation can be viewed as the extension of the 
pure heat transport process, since it takes into account the equations for momentum and heat fluxes:
\begin{eqnarray}
j_p &=& -L_{pp}\beta \frac{d \mu}{dy} + L_{pq} \frac{ d \beta}{dy} \nonumber \\
j_q &=& -L_{qp}\beta \frac{d \mu}{dy} + L_{qq} \frac{ d \beta}{dy}  \,\,\, .
\label{eq:onsager1}
\end{eqnarray}
where $L_{xx}$ are the entries of the symmetric Onsager matrix
(for pure heat transport the only nonzero entry is $L_{qq} = \kappa \,\, \beta^2$).
In coupled transport phenomena, these quantities play the role of generalized transport 
coefficients and, usually,  they are expected to depend on $\beta$ and $\mu$.

In the energy--representation, rather than referring to $j_q$, the energy flux $j_h$ 
is considered, whose corresponding  thermodynamic force is the gradient of 
$\mu\beta$. The coupled transport equations read
\begin{eqnarray}
j_p &=& -L'_{pp}\frac{d\beta \mu}{dy}   + L'_{ph} \frac{d \beta}{dy}  \nonumber \\
j_h &=&  -L'_{hp} \frac{d\beta \mu}{dy} + L'_{hh} \frac{ d \beta}{dy} \; ,
\label{eq:onsager2}
\end{eqnarray}
where $L'_{xx}$ is a new symmetric  Onsager matrix, whose entries depend in general on 
$\beta$ and $\mu$. In both representations the validity of the 
set of the linear response equations is conditioned to the existence of local thermodynamic 
equilibrium.

\begin{figure}
\begin{center}
\includegraphics[width=8cm,clip]{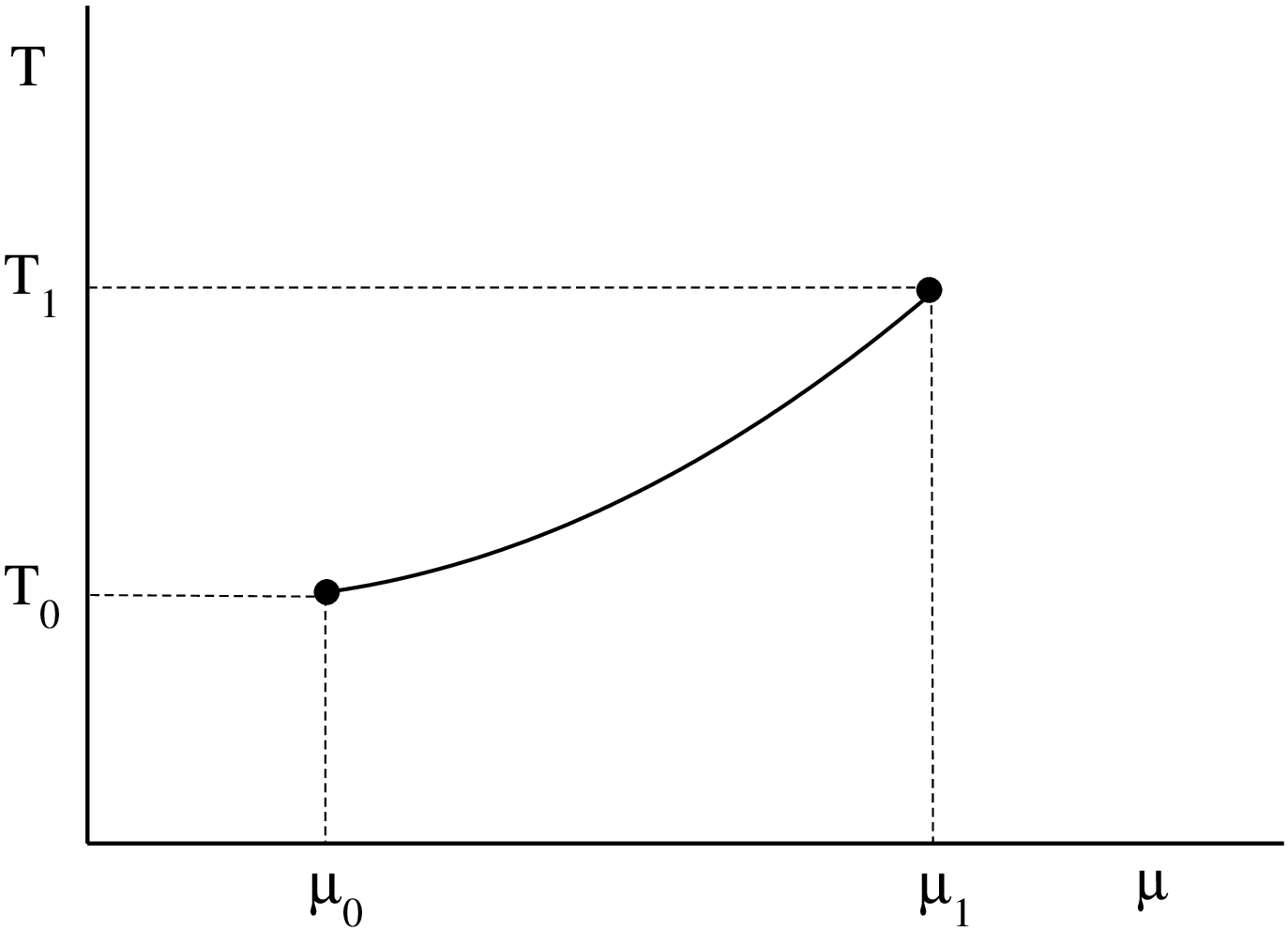}
\caption{Schematic view of the local equilibrium state of a chain in contact with two heat baths
at temperature $T_0$, $T_1$ and chemical potential $\mu_0$, $\mu_1$.}
\label{fig:general}
\end{center}
\end{figure}

As a suitable model for coupled transport in one dimension we consider a chain of particles, whose left ($y=0$)
and right ($y=1$) boundaries are
in contact with two reservoirs, operating at different temperatures, $T_0$ and $T_1$, 
and chemical potentials, $\mu_0$ and $\mu_1$. 
Within the $(\mu,T)$--plane, the variation of these thermodynamic variables along
the chain can be represented  as a path starting from an ``initial" state $(\mu_0,T_0)$ and ending in the 
``final" one $(\mu_1,T_1)$, or viceversa (see Fig.~\ref{fig:general}).  This task can be naturally accomplished  in
the energy-representation. In fact, when a stationary regime is established, $j_p$ and $j_h$  have to be  
constant along the chain. Accordingly, the shape of the path shown in Fig.~\ref{fig:general} is obtained by
integrating the set of differential equations (\ref{eq:onsager2}). There are two important remarks about the integration
procedure: $(i)$ it can be performed explicitly if the dependence of the corresponding
Onsager matrix elements $L'_{xx}$ on $\beta$ ($T$) and $\mu$  is known; 
$(ii)$ the set of differential equations have to fulfill  four boundary conditions, that 
fix the values of the two temperatures and  of the two chemical potentials imposed
by the reservoirs. These four conditions suffice to determine the values
of the two fluxes and of the two integration constants.

Notice that  when stationary conditions for coupled transport are established, 
$j_q$, at variance with $j_h$, is not constant along the chain and the path reconstruction 
in $(\mu,T)$--plane in the heat--representation is more involved.
This notwithstanding,  the set of equations (\ref{eq:onsager1})  reveals 
useful for studying coupled transport in models like the XY chain.
This is  a model of nearest-neighbour coupled rotors, whose interaction
energy depends on a phase variable $\phi$. The equations of motion
read
\begin{equation}
I \ddot \phi_i = U\left[ \sin(\phi_{i+1}-\phi_i) - \sin(\phi_i-\phi_{i-1}) \right]
\label{eq:rotors}
\end{equation}
where $I$ is the moment of inertia of the rotors, $U$ is the amplitude of the potential energy barrier
and the integer $i$ labels the sites along the chain ($y = i/N$ and $i= 0,\cdots,N$).

The microscopic expressions for momentum and heat fluxes of the model are  
\begin{equation}
\hspace{-1.5cm}  j_p(i) = -U\langle\sin(\phi_{i+1}-\phi_i) \rangle \quad , \quad
j_q(i) = -U\langle(\dot \phi_i-\langle \dot \phi_i\rangle)\sin(\phi_{i+1}-\phi_i) \rangle
\label{definitions}
\end{equation}
where the average $\langle \cdot \rangle$ is over stationary conditions
yielding local thermodynamic equilibrium. The chemical potential $\mu$ 
coincides with the rotation frequency $\omega$ of the rotors. The
term proportional to $\langle \dot \phi_i\rangle$ is precisely what 
we referred to above as the coherent part of the flux.

The dependence of the interaction term in the equation of motion
(\ref{eq:rotors}) on a trigonometric function of the phase variables
induces quite peculiar features of coupled transport.
In the heat representation, the off-diagonal terms $L_{qp}=L_{pq}$ vanish:
the heat current cannot induce a momentum current in a system which, 
on average, does not rotate. Therefore, Eq.~(\ref{eq:onsager1})  simplifies to
\begin{eqnarray}
j_p &=& -L_{pp}\beta \frac{d \omega}{dy} \nonumber \\
j_q &=&  L_{qq} \frac{ d \beta}{dy} \; .
\label{eq:onsager1a}
\end{eqnarray}
Moreover, $L_{pp}$ and $L_{qq}$ cannot depend on $\omega$. 
In fact, given any local oscillation frequency $\omega$, one can always choose a suitable rotating
frame where $\omega=0$. Since the physical properties of coupled transport must be independent
on the choice of the reference frame, $L_{pp}$ and $L_{qq}$ should depend on $T$ only.
At the first glance, these arguments seem to suggest that the underlying physics is 
pretty trivial, since it  corresponds to two uncoupled transport processes in the heat--representation. However, passing to the energy--representation, where
\begin{equation}
j_h = j_q + \omega j_p  \; ,
\label{eq:fluxh}
\end{equation}
simple calculations reveal that
\[ L'_{pp} = L_{pp} \quad , \quad L'_{ph}=L'_{hp} = L_{pp}\omega \quad , \quad 
L'_{hh} = L_{qq} + \omega^2L_{pp}  \; . \]
Altogether, the matrix $L'$ is symmetric (as it should) and, more importantly, 
its off-diagonal terms do not vanish. The relationship with the heat representation reveals
that the three coefficients defining $L'$ are not independent: all statistical properties
of the XY model are captured by two quantities only: $L_{pp}$ and $L_{qq}$.

In order to obtain a complete characterization
of coupled heat transport of the XY model in the energy--representation  
one has to determine the actual value to be attributed to $\omega$, 
since it depends on the rotating reference frame adopted for the entire system. 
Notice that this situation is analogous to the standard ambiguity of defining a potential
up to a constant or of  fixing a suitable gauge. 

This problem can be solved by shifting the origin of the frequency axis in such a way
that the energy flux vanishes. Once we have introduced 
\begin{equation}
\omega_e = \omega - \overline \omega \; ,
\label{eq:freqdef}
\end{equation}
the condition $j_q + \omega_e j_p = 0$ (see Eq.~(\ref{eq:fluxh})) implies
\begin{equation}
\overline \omega = j_h/j_p \; .
\end{equation}
As long as $j_p \ne 0$, $\overline \omega$ is a well defined variable.
Accordingly, we can ``fix the gauge"  by measuring the frequency $\omega$ in the reference 
frame where the energy flux vanishes. 

As a final step, we want to reconstruct the path described in the plane $(\omega,\beta)$, while
moving along the chain. It can be obtained by dividing 
term by term the two equations in (\ref{eq:onsager1a}) and by recalling that  
$j_q=-\omega j_p$ (see Eq.~(\ref{definitions})).
One finds the simple equation
\begin{equation}
\frac{d\beta}{d\omega} = \frac{L_{pp}}{L_{qq}}\beta\omega \; ,
\label{pathXY}
\end{equation}
where both $L_{pp}$ and $L_{qq}$ depend only on $\beta$.
It is convenient to rewrite Eq.~(\ref{pathXY}) in terms of the temperature $T$ and the squared
frequency  $\sigma = \omega^2$
\begin{equation}
\frac{d T}{d\sigma} = - \frac{L_{pp}}{L_{qq}} \frac{T}{2} \; .
\label{pathXYn}
\end{equation}
The path in the  $(\sigma,T)$--plane  can be obtained by formally integrating the
above equation 
\begin{equation}
\sigma = \int_T^{T_{max}} d\tau \frac{L_{qq}}{L_{pp}} \frac{2}{\tau}
\label{sigma}
\end{equation}
where $T_{max}$ is the maximum  value reached by the temperature $T$ along the chain
(see section \ref{numsim}).

\section{Thermal baths}
\label{thbath}
Various schemes can be employed for modeling the heat exchange of a physical system 
with a reservoir. The two most widely used are: (i)
Langevin heat baths; (ii) stochastic collisions \cite{LLP03,DHARREV}.
The former setup amounts to adding a pair of dissipating/fluctuating terms
to the equations of motion of the boundary particles.
In the latter one, the boundary particles are assumed to exchange
their velocity with equal-mass particles from an external heat bath, 
in equilibrium at some given temperature $T$.

Both schemes can be easily generalized to account for an exchange of
angular momentum, as well. In Ref. \cite{Iubini2014}, the following Langevin scheme was proposed (here 
we just refer to the last particle)
\begin{equation}
\hspace{-1.cm} I\ddot \phi_N = F(\phi_{N}-\phi_{N-1}) - F(\phi_{N+1}-\phi_{N}) 
 +\gamma(\omega_1 - \dot \phi) + \sqrt{2\gamma T}\,\xi(t) \; ,
 \label{Langevin}
\end{equation}
where the function $F$ is the torque acting between nearest--neighbour particles 
and $\omega_1$ can be interpreted as the frequency, or chemical potential, 
imposed by the stochastic bath via the external torque $\gamma \omega_1$,
where $\gamma$ defines the coupling strength with the bath. 
The quantity $\xi(t)$ accounts for a Gaussian white random noise 
with zero mean and unit variance, while the value of $\phi_{N+1}$ 
depends on the choice of boundary conditions: i.e. it is set equal to $\phi_N$ 
for open boundary conditions, or to 0 for fixed boundary conditions.

In the stochastic approach, the action of a reservoir imposing an average
frequency $\omega_1$ can be simulated by randomly resetting the velocity $\dot \phi_N$ 
of the end particle at random times (with some given average frequency), according
to the distribution
\[
P(\nu)  =   \sqrt{\frac{I}{\pi T}} \mathrm{e}^{-I(\nu - \omega_1)^2/{T}} \; .
\]
In this scheme, the bath frequency $\omega_1$ enters as a shift of the Gaussian distribution 
of $\nu$ \footnote{It is worth mentioning that a different strategy has been adopted by 
the authors of \cite{Pino2016}, who have explored a case where no heat exchange is involved. 
They have assume directly $\phi_{N+1} = \omega_1 t$ (without any extra torque).}.

\begin{figure}[ht]
\begin{center}
\includegraphics[height=0.5\textwidth,clip]{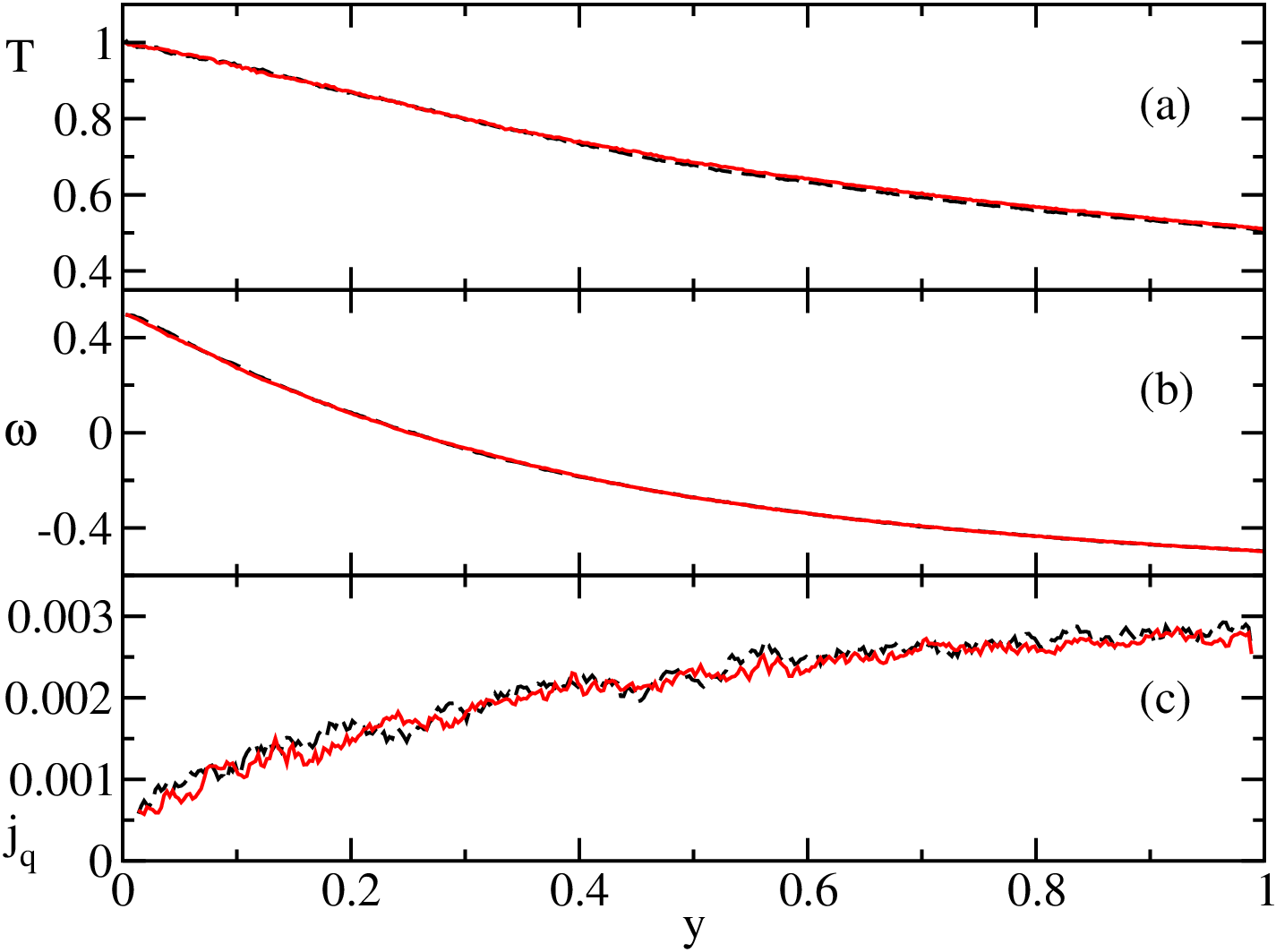}
\caption{Stationary nonequilibrium profiles corresponding to the parameters $\omega_0=0.5$,
$\omega_1=-0.5$, $T_0=1$, $T_1=0.5$ in a chain with $N=400$ particles.
(a) Temperature profiles; (b) frequency (chemical potential) profiles; (c) local heat fluxes.
Black dashed curves refer to Langevin heat baths with coupling parameter $\gamma=1$, implemented within a 4-th order Runge-Kutta integration scheme (time step $10^{-2}$). Red curves
are obtained using collisional heat baths with Poissonian distribution of collision times 
$\sim \exp(-\gamma_c t)$ and $\gamma_c=1$, implemented within a 4-th order MacLachlan-Atela (symplectic) integrator (time step $10^{-2}$ and total integration time $10^7$).}
\label{fig:cfrBATHS}
\end{center}
\end{figure}

Numerical tests reveal that these schemes are essentially equivalent to one another at
finite temperature (see the simulation data reported in Fig.~\ref{fig:cfrBATHS}).
However, this equivalence does not hold anymore in the limit case
where temperature is set to 0. Indeed there is a difference 
in the two schemes: the collisional setup maintains some
stochasticity due to the random times of the collisions, while the
Langevin setup reduces to a purely deterministic (dissipative) dynamics.
The interesting consequences emerging from such a difference will be investigated in a separate 
paper, devoted to a specific study of the limit case of zero-temperature heat baths.

\section{Numerical simulations of coupled transport in the XY chain}
\label{numsim}

\begin{figure}[ht]
\begin{center}
\includegraphics[height=0.5\textwidth,clip]{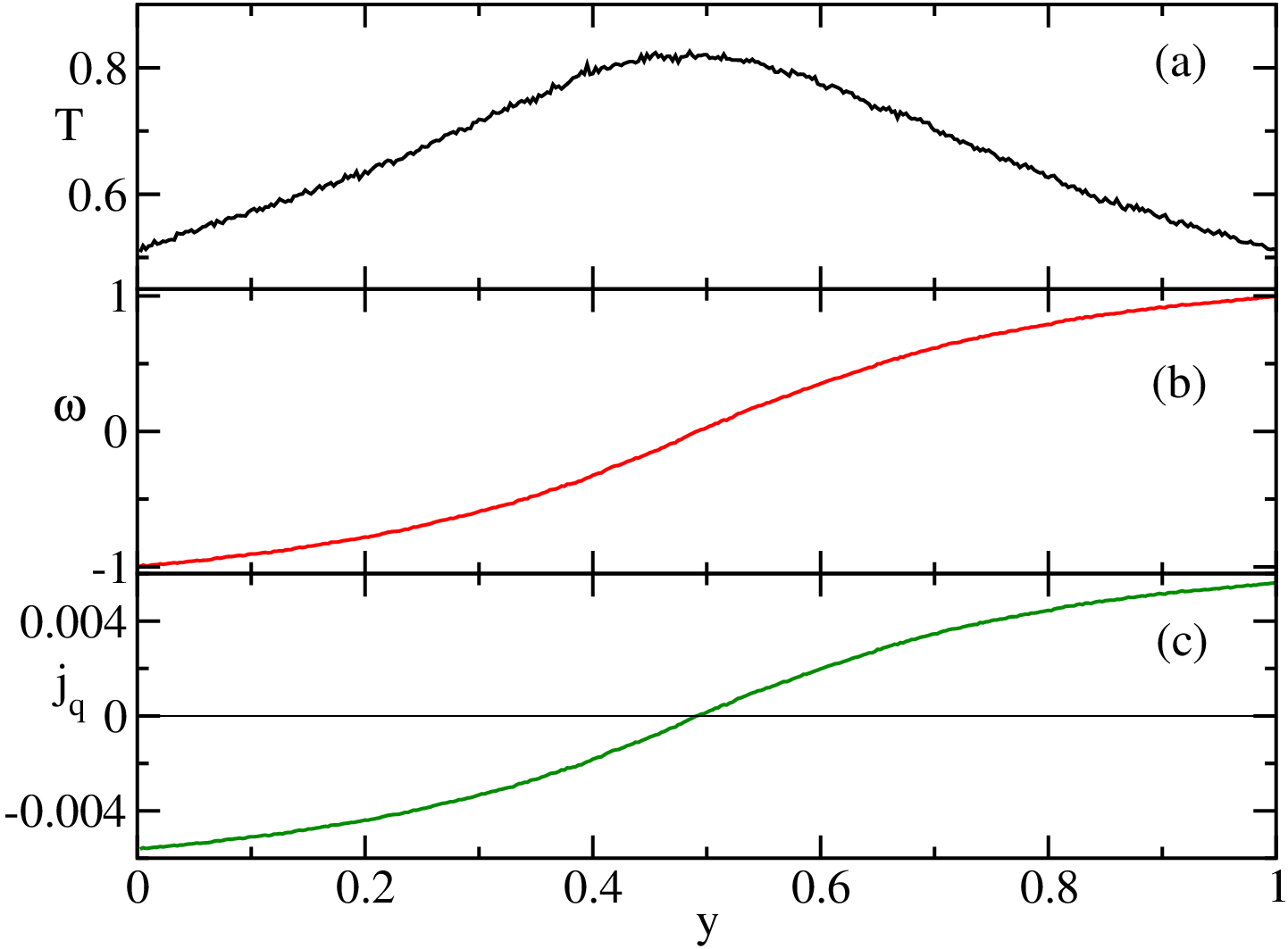}
\caption{Some observables for an XY chain of 400 particles, in contact at its 
boundaries with two
heat baths  at
temperature $T_0 = T_1 =  0.5$ and in the presence of  torques $\omega_0=-1$ and
$\omega_1=1$: (a) temperature
profile; (b) frequency (chemical potential) profile; (c)  local heat flux.
}
\label{fig:profile}
\end{center}
\end{figure}

We start this section by illustrating qualitatively how coupled transport manifests itself. 
In Fig.~\ref{fig:profile}, we show the frequency and temperature profile in
a case where both thermal baths are set to the same temperature and torques
$\omega_0 = -1$ and $\omega_1=1$ are applied at the chain ends.

The temperature profile exhibits a bump in the middle of the chain (as first 
found in \cite{Iacobucci2011}). The variation of the temperature along the chain
is a consequence of the coupling with the momentum flux imposed by 
the torque at the boundaries, although, in the end, the energy flux vanishes
(for symmetry reasons).
By recalling that $j_h = j_q + \omega j_p$ we see that the heat flux
$j_q = - \omega j_p$ varies along the chain being everywhere proportional
to the frequency, so that it is negative in the left part and positive in
the right side (this is again consistent with symmetry considerations).
In practice one can conclude that heat is generated in the central part,
where the temperature is higher and transported towards the two
edges. The total energy flux is however everywhere zero as the heat flux
is compensated by an opposite coherent flux due to momentum transfer.

Altogether, the presence of the temperature bump can be interpreted as a sort of Joule effect: the transport
of momentum involves a dissipation which in turn contributes to increasing
the temperature, analogously to what happens when an electric wire is crossed by a flux
of charges.

The flux of momentum $j_p$ is also obviously constant along the chain.
It can be used to determine the 
dependence of $L_{pp}$ on $T$,
\begin{equation}
L_{pp} = T j_p \frac{d y}{d \omega} \,\,\, .
\label{eq:oo3}
\end{equation}
Since only the differential of $\omega$ is involved in this equation, there is no need to distinguish
between $\omega$ and $\omega_e$ (see Eq.~(\ref{eq:freqdef})).

\begin{figure}[ht]
\begin{center}
\includegraphics[width=0.8\textwidth,clip]{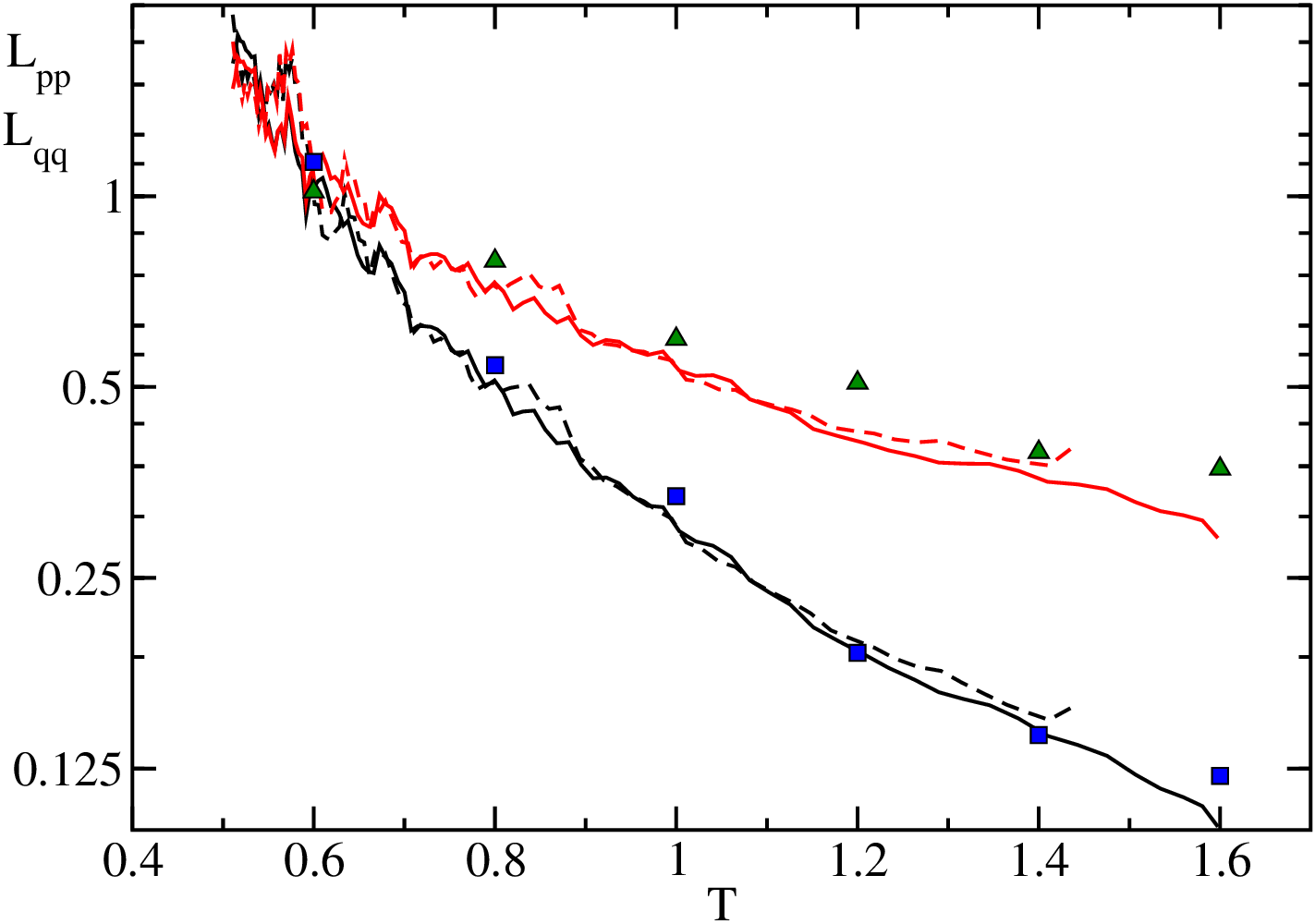}
\caption{
The two diagonal coefficients of the Onsager matrix obtained using Eq.~(\ref{eq:oo3})
in simulations of a chain 800 sites long and Langevin dynamics Eq.~(\ref{Langevin}). 
All curves correspond to $T_1=0.5$. The solid curves have been obtained for $\omega_1=2$,
while at $y=0$ no thermal bath nor torque and fixed boundary conditions have been applied. 
The dashed lines correspond to $\omega_1=1.5$, $\omega_0=0$ and $T_0=1.5$. 
Black (lower) and red (upper) lines correspond to $L_{pp}$ and $L_{qq}$, respectively.
The symbols have been obtained by implementing the Langevin reservoirs, Eq.~(\ref{Langevin})
to impose either small differences of temperature or chemical potential and thereby invoking
Eq.~(\ref{eq:onsager1a}).  More precisely, $L_{qq}$ (green triangles) was computed imposing a
temperature gradient $\Delta T=T_1-T_0=2T/5$ and $\Delta \omega=\omega_1-\omega_0=0$, while
$L_{pp}$ (blue squares) was obtained by setting $\Delta T=0$ and $\Delta \omega=T/4$.
Simulations refer to an XY chain with $N=512$.
}
\label{fig:lpp}
\end{center}
\end{figure}

\begin{figure}[ht]
\begin{center}
\includegraphics[width=0.8\textwidth,clip]{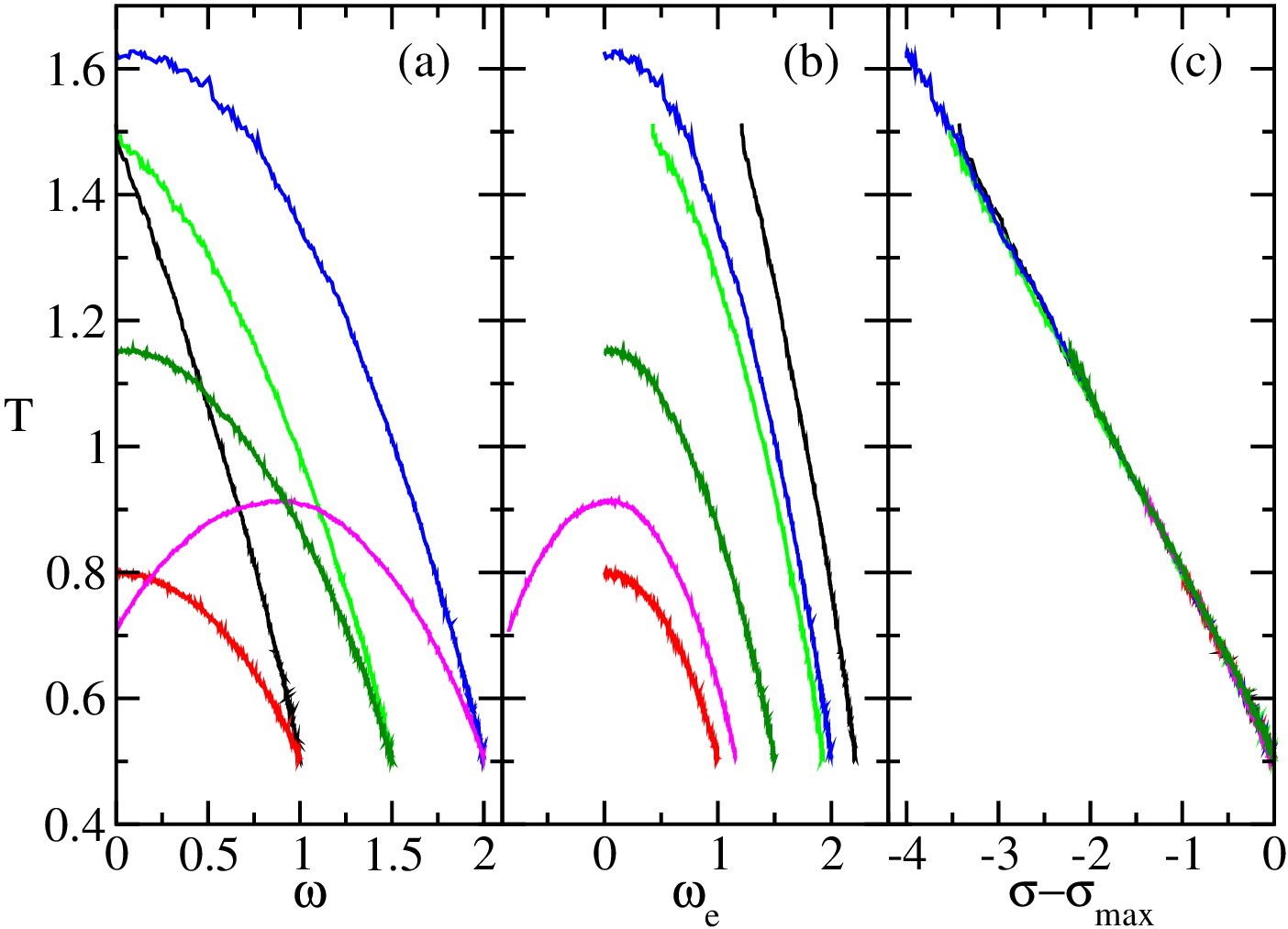}
\caption{Different representations of various nonequilibrium paths (the vertical scale
is logarithmic). Panel $(a)$ refers
to the real frequency $\omega$, observed in the numerical experiments; the frequency in panel
$(b)$ is shifted as in Eq.~(\ref{eq:freqdef})), i.e. it corresponds to the frequency 
measured in the frame, where the energy current vanishes; panel $c$ refers to the square
square effective frequency $\sigma=\omega_e^2$, suitably shifted to let the curves 
start from the same point in the bottom right.
All curves refer to an XY chain of 800 particles. In all simulations, fixed boundary 
conditions are assumed on the left boundary and a temperature $T_1=0.5$ is fixed on the right.
The red, dark green, and blue lines have been obtained with no heat bath on the left
and $\omega_1=$ 1, 1.5, and 2, respectively.
The purple line corresponds to $T_0=0.7$, $\omega_1=2$, the black line corresponds to
$T_0=1.5$, $\omega_1=1$, the light green line corresponds to $T_0=1.5$, $\omega_1=1.5$.
}
\label{fig:profiles3}
\end{center}
\end{figure}

The results of numerical simulations are plotted in Fig.~\ref{fig:lpp}:
the black curves are obtained by simulating a long chain submitted to a
relatively large temperature difference, while the squares correspond
to small gradients. The good agreement confirms the assumption of
a local thermal equilibrium.
$L_{pp}$ exhibits a  divergence for decreasing values of $T$ (notice that
the vertical axis is logarithmic). This reflects
the Arrhenius-type behavior of the thermal conductivity that has been previously
demonstrated \cite{Giardina99,Gendelman2000}.  
The red curves refer to $L_{qq}$: they have been obtained indirectly from the
knowledge of the ratio $L_{pp}/L_{qq}$, determined by following the procedure
described here below.

We started performing several sets of simulations. In all cases, we have imposed fixed boundary
conditions on the left side to ensure a zero frequency~\footnote{The frequency can be afterwards shifted 
by an arbitrary amount, without altering the physical properties}
and free boundary conditions on the right side with different values of the torque $f_1$ (see the caption of
Fig.~\ref{fig:profiles3} for additional details).
In some cases (see the red, dark green and blue lines in Fig.~\ref{fig:profiles3}a)
no thermal bath was used on the left boundary, which automatically implies a
vanishing heat flux (while the momentum flux self-adjusts on the basis of both boundary conditions). 

According to the theoretical considerations reported in Sec.~\ref{theory}, a meaningful
comparison among the different cases can be performed only after choosing a suitable
reference frame where the energy flux vanishes, i.e. by replacing $\omega$ with
the shifted frequency $\omega_e$  (see Eq.~(\ref{eq:freqdef})). The corresponding curves are 
reported in panel $(b)$ of Fig.~\ref{fig:profiles3}: the seemingly unphysical phenomenon
of mutual crossing of the different paths present in panel $(a)$ has disappeared, thus confirming
that $T$ and $\omega_e$ are proper thermodynamic variables. 
The final step of this data analysis consists in redrawing the paths in the $(\sigma,T)$--plane, where $\sigma=\omega_e^2$.
The result is shown in Fig.~\ref{fig:profiles3}c, where the abscissa has been chosen in such a way 
that all paths have in common the point corresponding to $T_1 = 0.5$
(notice that  $\sigma_{max} = (f_1 - \omega_e)^2$ is a path dependent quantity),
while the leftmost value of the abscissa for each path  corresponds to the maximum value 
of $T$ along the path.

The very good data collapse confirms that the 
thermodynamic behavior of coupled transport in the XY chain is determined by 
$T$ only. Notice that this result holds also for the purple path in Fig.~\ref{fig:profiles3}, 
which extends to negative values of  $\omega_e$: in fact, 
what matters is $\omega_e^2$, irrespective of the sign of the frequency itself.

From this analysis one understands that the origin of the temperature bump can be traced back
to a constant negative derivative of $dT/d\sigma$, which in turn follows from the positive sign
of the ratio $L_{pp}/L_{qq}$ which is fixed by thermodynamic conditions (see Eq.~(\ref{pathXYn})).

Moreover, the clean linear dependence of $T$ on $\sigma$ ($dT/d\sigma \approx - 0.28$) 
indicates that, at least in some temperature range\footnote{Preliminary simulations performed at smaller
temperatures suggest that the paths in the $(\sigma,T)$--plane bends down, while approaching the zero
temperature axis.}
Eq.~(\ref{pathXYn}) thus becomes
\begin{equation}
\frac{L_{qq}}{L_{pp}} \equiv D \approx \frac{T}{0.56I} \; .
\label{eq:XYLqqLpp}
\end{equation}
Here, the (equal to 1) moment of inertia $I$ has been added for dimensional reasons (the ratio
$L_{qq}/L_{pp}$ has the dimension of a squared frequency), to stress that 0.56 is a pure
adimensional number. We have no arguments to justify its value.  

By then making use of Eq.~(\ref{eq:oo3}), one can determine the dependence of $L_{qq}$ on $T$:
see the red lines displayed in Fig.~\ref{fig:lpp}. 
One could obtain $L_{qq}$ from from standard heat-transport simulations in the absence of momentum flux. 
The implementation of such a direct procedure to chains with small temperature gradient yields the 
green triangles reported in the same Fig.~\ref{fig:lpp}. The relatively good agreement confirms the
correctness of our approach to coupled transport.

\section{Coupled transport in the DNLS equation}
\label{dnls}
In this section we discuss coupled transport in the DNLS equation,
a more general model, where thermodynamic properties do not only depend
on the temperature, but also on the chemical potential. The evolution equation is
\begin{equation}
i \dot {z}_n = -2 |z_n|^2z_n - z_{n+1}-z_{n-1} \; ,
\end{equation}
where $z_n$ is a complex variable and $|z_n|^2$ is the local norm.
This system is particularly interesting because of its important applications in many domains of
physics ranging from waveguide optics, biomolecules and trapped cold gases \cite{Kevrekidis}.
The DNLS Hamiltonian has two conserved quantities, the mass/norm density $a$ and the energy 
density $h$ (for details see \cite{Rasmussen2000,Iubini2013}). Accordingly, it is a natural 
candidate for describing coupled transport \cite{Iubini2012,Iubini2013a}, which can be studied
by introducing the Langevin equation~\cite{Iubini2013a} (specified for the last lattice site)
\begin{equation}
\hspace{-1.cm} i \dot z_N= (1+i \gamma)\left[-2|z_N|^2 z_N  -z_{N+1}-z_{N-1} \right]  
 +i\gamma \mu z_N+ \sqrt{\gamma T} \, \eta(t) \quad  \; .
\label{ollac}
\end{equation} 
Here $\mu$ is the chemical potential imposed by the bath and $\eta(t)$ is a complex Gaussian white
noise with zero mean and unit variance.
In the high-temperature regime transport is normal \cite{Iubini2012}
and fluctuations of conserved fields spread diffusively \cite{Mendl2015}. 
However, in the low temperature regime phase slips are rare, with the consequence that phase 
differences appear as an additional (almost) conserved field, yielding anomalous transport on very 
long timescales \cite{Mendl2015}.

Unlike the XY model, the two currents associated with the conservation laws are mutually coupled 
in the DNLS equation. On the other hand, in a recent paper~\cite{Iubini2013a} 
it was argued that in the high mass-density limit (i.e. for large chemical potentials $\mu$)
the DNLS dynamics is well approximated by that of a XY chain in equilibrium simulations.
However, a precise identification of the parameter region where an accurate mapping is expected
has not yet been fully worked out. One of the reasons is the non uniformity of the
thermodynamic limit: no matter how long the system is, intermittent bursts always occur 
possibly invalidating the existence of a precise relationship.
It is therefore important to explore the connection between the two
models from the point of view of irreversible thermodynamics, comparing for instance the
associated Onsager coefficients.

In Ref.~\cite{Iubini2013a} it was found that in the large mass limit a thermostatted DNLS equation with parameters 
$T$ and $\mu$ is equivalent to the XY model,
\begin{eqnarray}
\label{XY+bath}
\hspace{-1.5cm} \dot{\phi_N} &=&p_N\\
\hspace{-1.5cm} \dot{p_N} &=& U \left[\sin(\phi_{N+1}-\phi_N) - \sin(\phi_{N}-\phi_{N-1})\right] -
\gamma'\left(p_N-\delta\mu\right) +
\sqrt{4\gamma' T} \, \xi(t)\nonumber \quad ,
\end{eqnarray}
where $\gamma'=U\gamma$. Here we have defined $\mu=(U/2-2)+\delta\mu$, which corresponds to describing the DNLS model 
in a rotating reference frame  with frequency $(\mu-\delta\mu)=(U/2-2)\gg\delta\mu$. 
This choice does not limit the generality of the mapping, since any other choice
of the reference frame would produce a shift of all the XY phase velocities that can be eliminated by the gauge 
transformation described in Section 2. Finally, by looking at the stochastic term and comparing it with the
analogous term in Eq.~(\ref{Langevin}) one notices a factor 2 difference in the definition of the temperature:
this point will be important later on.

From a thermodynamic point of view, the major difference between the rotor model and DNLS equation is that in the former
case, the off-diagonal elements $L_{pq}=L_{qp}$ vanish. Therefore, the adimensional Seebeck coefficient
\begin{equation}
 S=\frac{1}{T} \frac{L_{pq}}{L_{pp}}
\end{equation}
is a proper indicator to test the closeness of the two models.

\begin{figure}[ht]
\begin{center}
\hfill
\includegraphics[width=0.8\textwidth,clip]{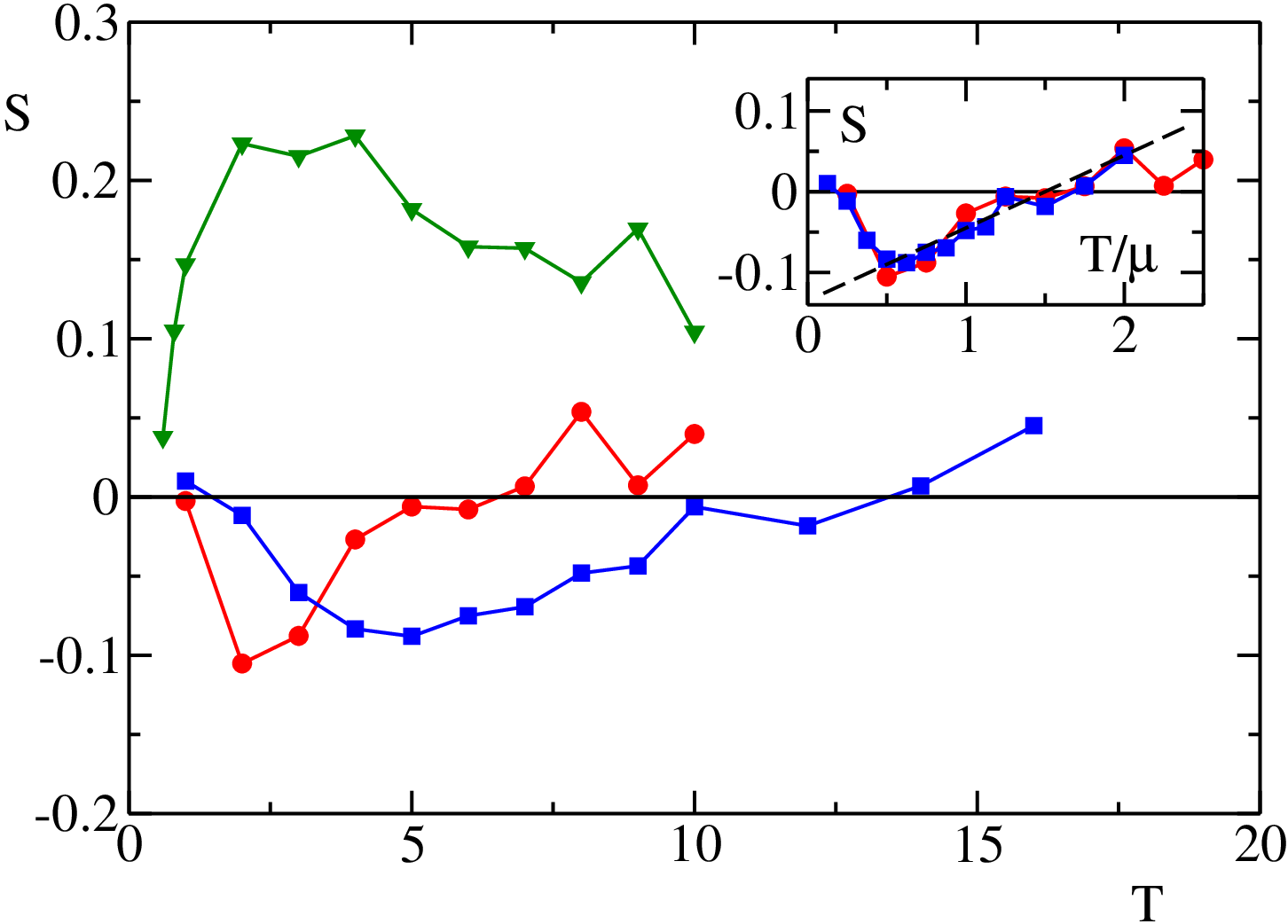}
\caption{The DNLS Seebeck coefficient $S$ as a function of the temperature $T$ for three different 
chemical potentials: $\mu=0$ (green triangles), $\mu=4$ (red circles) and $\mu=8$ (blue squares).   
In the inset: the curves with $\mu=4$ and $\mu=8$ in the  rescaled temperature  representation $T/\mu$.
The black dashed curve is a linear approximation of $S$ around $S=0$ with slope $0.09$.
Simulations have been performed imposing $\Delta T=0.1 T$ and $\Delta \mu=0.1$ on DNLS chains
with $N=400$ lattice sites initially in equilibrium at $T$, $\mu$. The three leftmost points of the curve
$\mu=0$ refer to DNLS with $N=800$.}
\label{fig:Seeb}
\end{center}
\end{figure}

Fig. \ref{fig:Seeb} shows the dependence of the Seebeck coefficient in the DNLS model on the temperature for
three different values of the chemical potential $\mu$. Upon increasing $\mu$ we indeed see that $S$ decreases
and crosses the zero axis for some finite temperature.
The two curves for $\mu=4$ and $\mu=8$  indicate that the zero-Seebeck condition occurs 
approximately for a temperature that is proportional to $\mu$,  $T_c \sim 1.5 \mu$ (see the inset). 
In the neighborhood of $T_c$ the Seebeck coefficient grows almost linearly $S\sim 0.09\,T/\mu$ (see the red dashed line). 
A consistent equivalence with the XY chain in a broad range of parameter values would require that upon increasing 
$\mu$ the slope should decrease. In so far as it stays constant, as our simulations seem to suggest, a 
quantitative agreement is restricted to a tiny temperature-interval around $T_c$. 

\begin{figure}[ht]
\hfill \includegraphics[width=0.8\textwidth,clip]{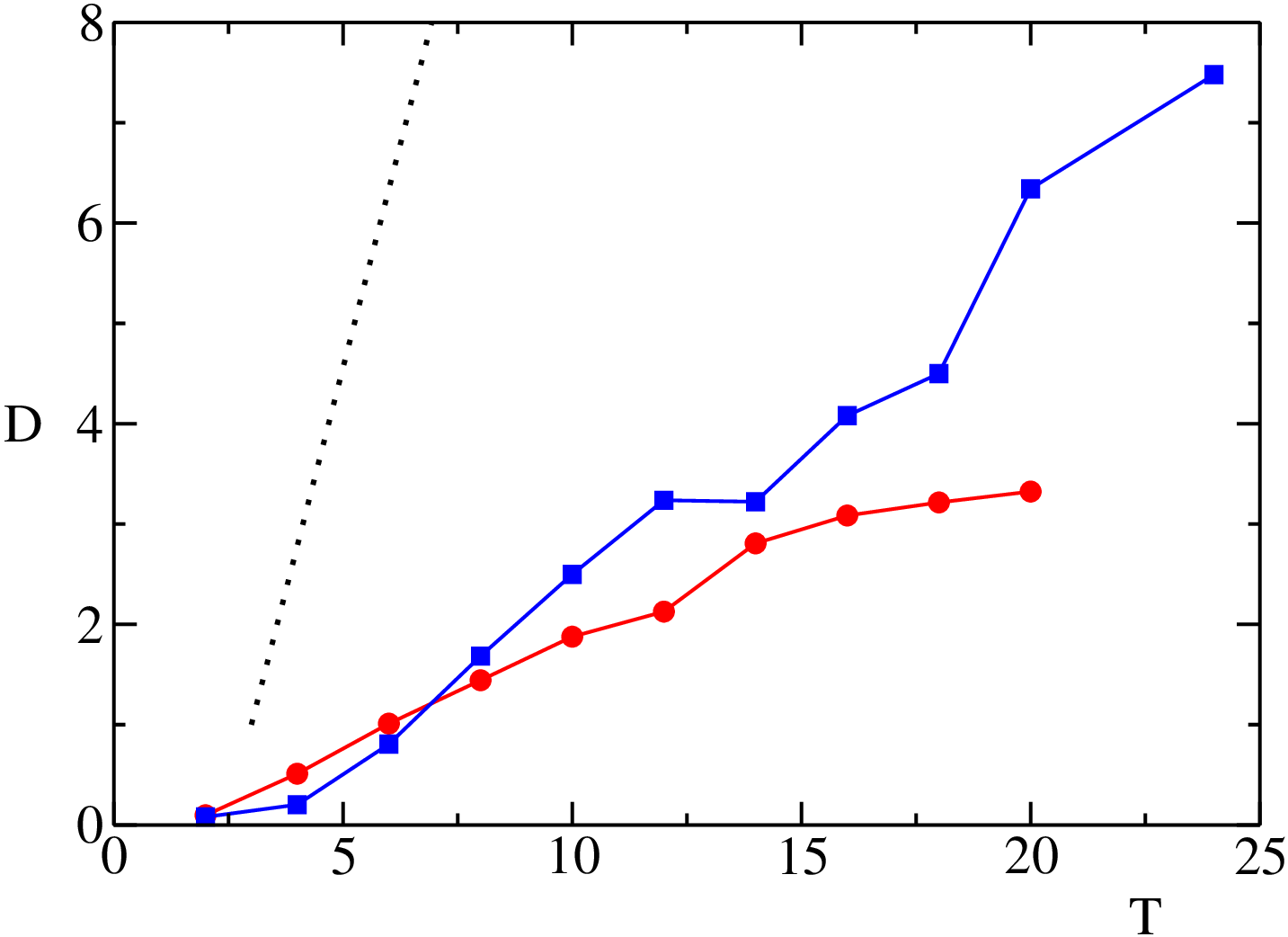}
\caption{The ratio of the diagonal Onsager coefficients $D=Lqq/Lpp$ for the DNLS equation with average chemical potential 
$\mu=4$ (red circles) and $\mu=8$ (blue squares). The simulation parameters are the same as in Fig.~\ref{fig:Seeb}.
The black dotted line corresponds to the slope as expected from the study of the rotor model.}
\label{fig:Lqq_over_Lpp}
\end{figure}

For a complete characterization of the DNLS transport, it is instructive to look also at the diagonal elements
of the Onsager matrix and, in particular at the ratio $D=L_{qq}/L_{pp}$.
In Fig.~\ref{fig:Lqq_over_Lpp} we plot $D$ as a function of the temperature $T$, multiplied by a factor 2, to take
into account the scale difference with the XY model. An approximately linear growth is found that is
analogous to the behavior observed in the rotor model. The slope is, however, smaller (see the dotted curve) although it keeps 
increasing with the value of the chemical potential. Accordingly we can conjecture that upon further increasing $\mu$
a better agreement could be found, but more refined simulations are necessary for a more quantitative statement.

\section{Discussion and conclusions}
\label{conclu}

In this paper we have provided a detailed analysis of the structure of
nonequilibrium steady states in the presence of coupled transport.
In both models (XY and DNLS), the relevant variable is an angle and that is the
reason why (especially in the XY setup) the coupling between angular momentum
and energy gives rise to nontrivial phenomena.
In generic nonlinear chains, like the Fermi-Pasta-Ulam model or similar
\cite{LLP03,DHARREV}, particles characterized by different velocities
would inevitably fall apart with no mutual interactions; in our set-up the
very nature of the angular variables (defined, say, between $0$ and $2\pi$)
induces a different physical scenario.
The XY-dynamics is nevertheless reminiscent of the evolution of nonlinear oscillators
in that no coupling is present between heat and angular momentum current
(this statement is equivalent to saying that the Seebeck coefficient is identically
equal to zero).
In spite of its extremely simplified structure, we have shown that this setup
can sustain coupled transport whenever a torque is applied to the chain ends. The reason
is due to the emergence of a coherent energy flux, which acts as a mediator.
It is foreseeable that a deeper understanding will be useful in the 
problem of nano and mesoscale heat transport. For instance, in the context 
of Josephson physics it has been recently demonstrated that some form of 
coherent heat transport may be used for control in special applications \cite{Giazotto2012}.     
As a coherent contribution is expected to arise in more general physical setups,
an important advice can be given for future studies, namely that of
singling it out and distinguishing it from the coupling which involves the heat
flux.

The DNLS is a model where heat flux is directly coupled with norm flux.
However, consistently with a previous claim~\cite{Iubini2013a}, our numerical simulations show that
in the limit of large chemical potential the DNLS equation reduces to the
XY rotor model. In particular, we find that the critical line separating
positive from negative values of the Seebeck coefficients, extend to large
$\mu$-values. This encourages the performance of further studies to put
the equivalence on a firmer basis and to possibly use the equivalence
as a starting point for a perturbative analysis.

Finally, additional studies of the XY model are welcome both in the region of small 
temperatures, where the confinement within the energy valley becomes crucial (we are currently working in
this direction) and of high-temperatures, where for different reasons
a dynamical ergodicity breaking is expected (see e.g. \cite{Pino2016}), which 
strongly modifies transport properties.

\section*{Acknowledgement}
One of us (AP) wishes to acknowledge S. Flach for enlightening discussions 
about the relationship between the DNLS equation
and the rotor model.

\section*{References}
\bibliography{diodo,heat}

\providecommand{\newblock}{}
\begin{thebibliography}{10}
\expandafter\ifx\csname url\endcsname\relax
  \def\url#1{{\tt #1}}\fi
\expandafter\ifx\csname urlprefix\endcsname\relax\def\urlprefix{URL }\fi
\providecommand{\eprint}[2][]{\url{#2}}

\bibitem{LLP03}
Lepri S, Livi R and Politi A 2003 {\em Phys. Rep.\/} {\bf 377} 1

\bibitem{DHARREV}
Dhar A 2008 {\em Adv. Phys.\/} {\bf 57} 457--537

\bibitem{basile08}
Basile G, Delfini L, Lepri S, Livi R, Olla S and Politi A 2007 {\em Eur. Phys
  J.-Special Topics\/} {\bf 151} 85--93

\bibitem{Lepri2016}
Lepri S (ed) 2016 {\em Thermal transport in low dimensions: from statistical
  physics to nanoscale heat transfer\/} ({\em Lect. Notes Phys\/} vol 921)
  (Springer-Verlag, Berlin Heidelberg)

\bibitem{Wang2013}
Wang L, Hu B and Li B 2013 {\em Phys. Rev. E\/} {\bf 88}(5) 052112

\bibitem{Wang2011}
Wang L and Wang T 2011 {\em EPL (Europhysics Letters)\/} {\bf 93} 54002

\bibitem{Lepri2009}
Lepri S, Mej{\'\i}a-Monasterio C and Politi A 2009 {\em J. Phys. A: Math.
  Theor.\/} {\bf 42} 025001

\bibitem{Das2014a}
Das S~G, Dhar A, Saito K, Mendl C~B and Spohn H 2014 {\em Phys. Rev. E\/} {\bf
  90} 012124

\bibitem{Lee-Dadswell2015}
Lee-Dadswell G 2015 {\em Phys. Rev. E\/} {\bf 91} 032102

\bibitem{Spohn2014}
Spohn H 2014 {\em J. Stat. Phys.\/} {\bf 154} 1191--1227

\bibitem{Das2014}
Das S, Dhar A and Narayan O 2014 {\em Journal of Statistical Physics\/} {\bf
  154} 204--213 ISSN 0022-4715

\bibitem{Liu2014}
Liu S, H{\"a}nggi P, Li N, Ren J and Li B 2014 {\em Phys. Rev. Lett.\/} {\bf
  112} 040601

\bibitem{Mejia2001}
Mej\'ia-Monasterio C, Larralde H and Leyvraz F 2001 {\em Phys. Rev. Lett.\/}
  {\bf 86} 5417--5420

\bibitem{Casati2009}
Casati G, Wang L and Prosen T 2009 {\em J. Stat. Mech.: Theory and
  Experiment\/}  L03004

\bibitem{Benenti2014}
Benenti G, Casati G and Mej{\'\i}a-Monasterio C 2014 {\em New J. Phys.\/} {\bf
  16} 015014

\bibitem{Gillan85}
Gillan M and Holloway R 1985 {\em J. Phys. C\/} {\bf 18} 5705--5720

\bibitem{Basko2011}
Basko D 2011 {\em Annals of Physics\/} {\bf 326} 1577 -- 1655

\bibitem{Iubini2012}
Iubini S, Lepri S and Politi A 2012 {\em Phys. Rev. E\/} {\bf 86} 011108

\bibitem{DeRoeck2013}
De~Roeck W and Huveneers F 2015 {\em Commun. Pure Appl. Math.\/} {\bf 68}
  1532--1568

\bibitem{Borlenghi2015}
Borlenghi S, Iubini S, Lepri S, Chico J, Bergqvist L, Delin A and Fransson J
  2015 {\em Phys. Rev. E\/} {\bf 92}(1) 012116

\bibitem{Escande1994}
Escande D, Kantz H, Livi R and Ruffo S 1994 {\em J. Stat. Phys.\/} {\bf 76}
  605--626

\bibitem{Giardina99}
Giardin{\'a} C, Livi R, Politi A and Vassalli M 2000 {\em Phys. Rev. Lett.\/}
  {\bf 84} 2144--2147

\bibitem{Gendelman2000}
Gendelman O~V and Savin A~V 2000 {\em Phys. Rev. Lett.\/} {\bf 84} 2381--2384

\bibitem{Yang2005}
Yang L and Hu B 2005 {\em Phys. Rev. Lett.\/} {\bf 94}(21) 219404

\bibitem{Li2015}
Li Y, Liu S, Li N, H{\"a}nggi P and Li B 2015 {\em New J. Phys.\/} {\bf 17}
  043064

\bibitem{Mendl2013}
Mendl C~B and Spohn H 2013 {\em Phys. Rev. Lett.\/} {\bf 111}(23) 230601

\bibitem{Das2014b}
Das S~G and Dhar A 2014 {\em arXiv preprint arXiv:1411.5247\/}

\bibitem{Flach2003}
Flach S, Miroshnichenko A and Fistul M 2003 {\em Chaos\/} {\bf 13} 596--609

\bibitem{Eleftheriou2005}
Eleftheriou M, Lepri S, Livi R and Piazza F 2005 {\em Physica D: Nonlinear
  Phenomena\/} {\bf 204} 230--239

\bibitem{Cuneo2015}
Cuneo N and Eckmann J~P 2015 {\em arXiv preprint arXiv:1504.04964\/}

\bibitem{Gendelman2010}
Gendelman O and Savin A 2010 {\em Phys. Rev. E\/} {\bf 81} 020103

\bibitem{Iacobucci2011}
Iacobucci A, Legoll F, Olla S and Stoltz G 2011 {\em Phys. Rev. E\/} {\bf 84}
  061108

\bibitem{Iubini2014}
Iubini S, Lepri S, Livi R and Politi A 2014 {\em Phys. Rev. Lett.\/} {\bf 112}
  134101

\bibitem{Ke2014}
Ke P and Zheng Z~G 2014 {\em Frontiers of Physics\/} {\bf 9} 511--518

\bibitem{Delfini2005}
Delfini L, Lepri S and Livi R 2005 {\em J. Stat. Mech: Theory Exp.\/}  P05006

\bibitem{Pino2016}
Pino M, Ioffe L~B and Altshuler B~L 2016 {\em Proceedings of the National
  Academy of Sciences\/} {\bf 113} 536--541

\bibitem{Iubini2013a}
Iubini S, Lepri S, Livi R and Politi A 2013 {\em J. Stat. Mech: Theory Exp.\/}
  P08017

\bibitem{Kevrekidis}
Kevrekidis P~G 2009 {\em The Discrete Nonlinear Schr\"odinger Equation\/}
  (Springer Verlag, Berlin)

\bibitem{Rasmussen2000}
Rasmussen K, Cretegny T, Kevrekidis P~G and Gr{\o}nbech-Jensen N 2000 {\em
  Phys. Rev. Lett.\/} {\bf 84} 3740--3743

\bibitem{Iubini2013}
Iubini S, Franzosi R, Livi R, Oppo G and Politi A 2013 {\em New J. Phys.\/}
  {\bf 15} 023032

\bibitem{Mendl2015}
Mendl C~B and Spohn H 2015 {\em J. Stat. Mech: Theory Exp.\/} {\bf 2015} P08028

\bibitem{Giazotto2012}
Giazotto F and Mart{\'\i}nez-P{\'e}rez M~J 2012 {\em Nature\/} {\bf 492}
  401--405

\end{thebibliography}
\bibliographystyle{iopart-num}

\end{document}